# Supercooled Droplet Icing and Self-Jumping on Micro/nanostructured Surfaces: Role of Vaporization Momentum

Xiao Yan[1=], Samuel C. Y. Au[1=], Sui Cheong Chan[1], Ying Lung Chan[1], Ngai Chun Leung[1], Wa Yat Wu[1], Dixon T. Sin[1], Guanlei Zhao[2], Casper H. Y. Chung[1], Mei Mei[1], Yinchuang Yang[1], Huihe Qiu[1], Shuhuai Yao[1]*

[1]Department of Mechanical and Aerospace Engineering, Hong Kong University of Science and Technology, Hong Kong, China

[2]State Key Laboratory of Automotive Safety and Energy, School of Vehicle and Mobility, Tsinghua University, Beijing 100084, China.

[=] *Equal contribution*
\* *Corresponding author email*: meshyao@ust.hk (Shuhuai Yao)

**ABSTRACT**

Phase change under reduced environmental pressures is key to understanding liquid discharge and propulsion processes for aerospace applications. A representative case is the sessile water droplets exposed to high vacuum, which experience complex phase change and transport phenomena that behave so differently than that under the atmosphere. Here, we demonstrate a previously unexplored aspect of the mechanism governing icing droplet self-launching from superhydrophobic surfaces when exposed to low pressures (~100 Pa). In contrast to the previously reported recalescence-induced local overpressure underneath the droplet that propels icing droplet self-jumping, we show that the progressive recalescence over the free surface plays a significant role in droplet icing and jumping. The joint contribution of the top-down vaporization momentum and bottom-up local overpressure momentum leads to vaporization-compression-detaching dynamics of the freezing droplets. We delineate the jumping velocity of the icing droplet by analyzing droplet vaporization mediated by freezing and substrate structuring, and reveal jumping direction coupled with the spatially probabilistic ice nucleation. Our study provides new insights into phase change of supercooled droplets at extreme conditions seen in aerospace and vacuum industries.

Water droplet freezing on solid surfaces poses safety and economic threats to transportation infrastructure, power generation/transmission systems, and telecommunication facilities.[1–4] A classic example in the aerospace industry is the in-flight icing of supercooled droplets that has caused >40% of general aviation accidents.[5,6] More recently, the global energy transition to renewable energies has exacerbated the issue of ice accretion on wind turbine blades and photovoltaic panels, which causes significant power loss and serious safety hazards.[7–10]

Icing of a droplet on solid surfaces starts from ice nucleation[12] followed by crystallization propagation accompanied by the release of latent heat (recalescence).[13] Extensive studies during the past decades have shed light on the icing physics governing sessile droplet icing,[14–17] condensation frosting,[18,19] and impact droplet icing.[20,21] The developed understanding, in turn, contributes to the development of various icephobic surfaces such as superhydrophobic/biphilic surfaces,[22,23] liquid or solid-state slippery surfaces,[24] and gel-type or polymer-based low-toughness surfaces[25–27] that can regulate ice accretion by removing supercooled droplets before freezing, delaying ice nucleation, and reducing ice adhesion.

Despite the rapid growth of knowledge in icing physics and icephobic surface design, most of the studies on icing confined themselves within atmospheric conditions with little attention being put on the environmental effects beyond temperature. Different from atmospheric icing, droplet icing in a low environment pressure has a higher vaporization flux over the droplet interface, leading to substantial supercooling that enables fast icing from the free surface of the droplet. Previous work has reported that droplet icing can behave so differently under a low-pressure environment,[13] showing that icing droplets can depart from the surface by vaporizing intensively beneath the droplet due to recalescence that creates an overpressure within the microstructures of superhydrophobic surfaces.[13] Besides, self-detachment of icing droplets can be achieved via preferential nucleation on the droplet's free surface that produces a concentric inward growth of freezing front, displacing the liquid within the ice shell and leading to a self-dislodging[28] and even explosion.[29]

Regardless of these remarkable observations of self-detachment of icing droplets under low pressures, the ice detaching mechanism remains to be further clarified. Specifically, the prevailing overpressure model[28] attributes the droplet jumping to the local vaporization facing the substrate at the end of icing while ignoring the vaporization accompanied by ice propagation. The latter has been shown to introduce intensive outward vapor flow,[30] which may also contribute to droplet detachment. Furthermore, the proposed overpressure model assumes a vaporization flux inversely dependent on the microstructures thus leading to an infinite overpressure when the structure scale reduces to infinitely small,[13] which is unphysical. More importantly, the icing-droplet detaching kinetics (velocity and direction), key to understanding the icing physics, remain to be investigated.

Here, we demonstrate an unexplored aspect of the mechanisms governing icing-droplet jumping on superhydrophobic surfaces under a rapidly depressurized environment. In addition to causing the previously reported substrate-mediated overpressure, we show that the icing of a supercooled droplet also leads to a strong vaporization momentum due to the progressive propagation of the icing front. The collaborative contribution of the bottom-up (overpressure) and top-down (progressive vaporization) is characterized by the compressive deformation of the icing droplet followed by icing propagation-dependent jumping. By performing droplet icing on superhydrophobic surfaces having differing structures and characterizing droplet jumping dynamics, we develop a model to incorporate the vaporization momentum, which allows us to quantify droplet jumping velocity and direction.

Figure 1a shows the experimental setup to study icing and departure dynamics of supercooled droplets at low pressures (see also Figure S1, Section S1 of Supporting Information for more details on experimental setup and conditions). A deionized water droplet (6-30 μL) was initially deposited on a horizontal superhydrophobic substrate. The superhydrophobic substrate has an apparent advancing contact angle $\theta_a^{app}$=163.2±1.8º and an apparent receding contact angle $\theta_r^{app}$=158.8±1.6º (see Figure S3, Section S2 of Supporting Information). After equilibrating at standard room temperatures and pressures (STP) for ≈2 min, the droplet and substrate were transferred to a customized chamber for depressurization tests. The vacuum chamber was coupled to a mechanical pump (Agilent DS 602 Rotary Vane Pump) to achieve fast depressurization from atmospheric pressure (101 kPa) to ~100 Pa within 10-20 s (Figure 1b). The depressurization rate is 0.1 bar/s, comparable with the previous study (see Figure S1b, Section S1 of Supporting Information).[13] High-speed optical (PCO.DIMAX CS3) and thermal (FLIR SC7700) imaging was performed from a side view and a top view to capture the droplet icing and jumping dynamics.

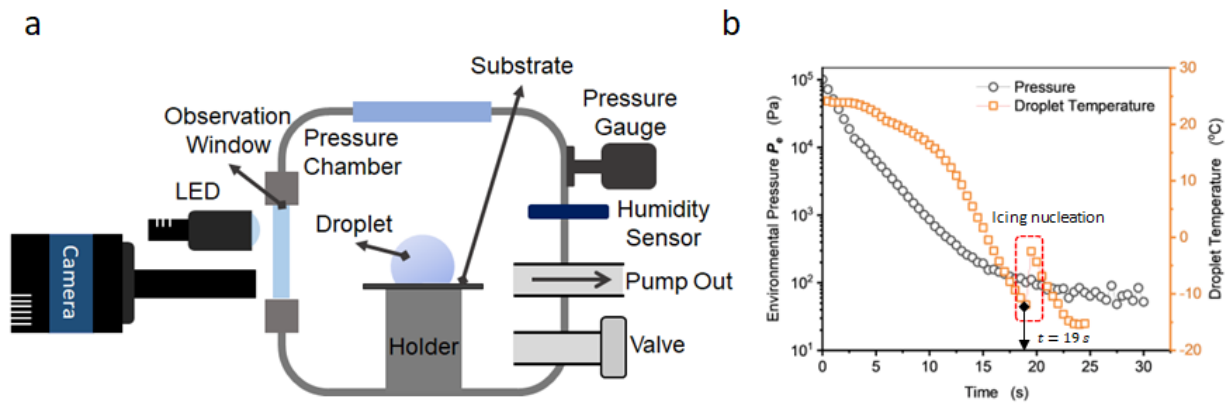

**Figure 1.** (a) Schematic of the experimental setup for the ice jumping visualization. The water droplet is placed on the substrate which is fixed on the substrate holder. The high-speed camera is used to capture the droplet icing dynamics. The environmental pressure is measured by the pressure gauge. The sapphire window allows for optical and thermal (mid-wave infrared, 1.5-5 μm) imaging. (b) Chamber pressure and droplet temperature as a function of time during chamber depressurization. Droplet temperature is measured by inserting a thermocouple into the droplet residing on the substrate. As the chamber pressure decreases from 101 kPa to ~100 Pa, the droplet is supercooled to reach a temperature of around -12ºC and then suddenly heats up to ≈0ºC, indicating the occurrence of freezing and recalescence (marked by red dotted frame). See Figures S1 and S2, Section S1, Supporting Information for details on experimental setup and procedures as well as thermodynamic conditions.

Due to the evaporative cooling at the liquid-gas interface (a heat flux up to 50 kW/m$^2$, see Figure S1d, Section S1 of Supporting Information), the droplet surface temperature decreased to a supercooled temperature (below -12 ºC, Figure 1b) until ice nucleation initiated from the free surface of the droplet (Figure 2a, see Video S1). Upon ice nucleation, the icing front (marked by the yellow boundary and line, $t$=2.6 ms in Figure 2a and $t$=5 ms in Figure 2b) propagated fast along the droplet surface. The icing was accompanied by the rapid release of latent heat of solidification, which led to recalescence and heating up of the ice-water slurry from supercooled to ≈ 0ºC (equilibrium freezing temperature) as indicated by the infrared (IR) imaging (Figure 2c, see Video S2). As the ice shell was formed to enclose the droplet, the droplet underwent compressive deformation from a sphere cap to an ellipse ($t$=6.73ms, Figure 2a), followed by the self-detachment of the solidifying droplet from the substrate ($t$=9.33ms, Figure 2a). In addition to out-of-plane jumping, the icing droplet also experienced in-plain movement depending on the progressive icing, as confirmed by the top-view visualization (Figure 2b, see Video S1).

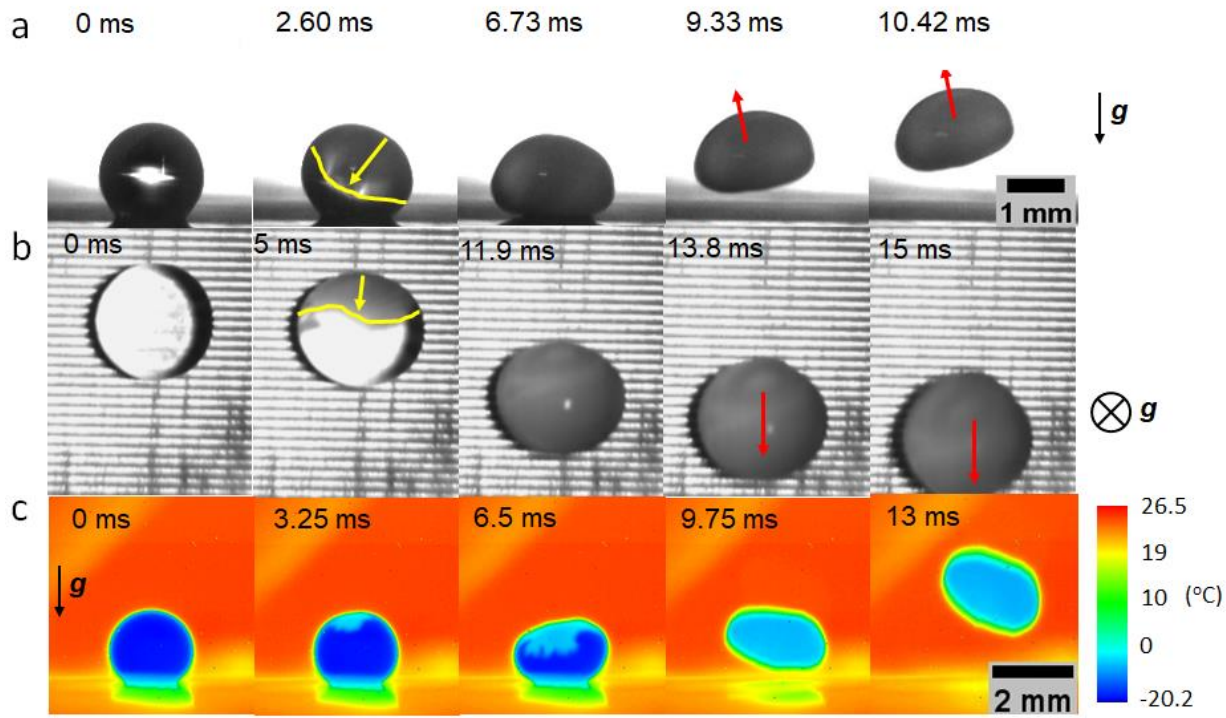

**Figure 2.** (a) Side-view and (b) top-view high-speed time-lapse images of a 10 μL water droplet solidifying and jumping on micro/nanostructured superhydrophobic surfaces upon depressurization. The yellow lines and arrows indicate the position of the freezing front and the freezing propagation direction, whereas the red arrows indicate the jumping direction of the icing droplet. (c) Side-view thermography of an icing and jumping droplet, showing the ice nucleation and propagation (light blue) over the free surface of the droplet (dark blue). The direction of gravity is indicated by the arrows in (a) and (c) or the circled cross in (b) (gravity pointing to the paper plane). See Videos S1 and S2.

In contrast to the previously reported icing droplet levitation induced by local overpressure underneath the droplet on microstructured superhydrophobic surfaces,[28] the observed droplet jumping here is characterized by the droplet compressive deformation and icing propagation-dependent jumping dynamics. We thus propose a physical model to account for the deformation and jumping of the icing droplet, i.e., the vaporization-deformation-bouncing model as illustrated in Figure 3. The primary assumption is the intensive release of vapor from the progressive recalescing surface, which leads to propulsion towards the ice shell, pushing the liquid to deform. The concentric and inward growth of the ice shell leaves the inner of the droplet partially unfrozen as the droplet deforms towards the substrate. Mediated by this unfrozen portion of the droplet interacting with the superhydrophobic substrate, the deformation momentum is accumulated and eventually redirected to propel out-of-plane droplet jumping, in a similar way to droplet impact and bouncing.[31]

To understand the intensive vaporization assumed in our physical model, we first quantified the vaporization flux resulting from recalescence during icing. The vaporization flux ($J$) is driven by vapor pressure difference, as revealed by the Schrage equation $J \sim \left(\frac{P_d}{\sqrt{T_d}} - \frac{P_e}{\sqrt{T_e}}\right)$,[32,33] where $P_d$ and $T_d$ are the local vapor pressure and temperature at the droplet interface (ice or water), respectively; $P_e$ and $T_e$ are the environmental pressure and temperature, respectively. Using $T_d$, $P_e$, and $T_e$ measured at the moment of ice nucleation, we demonstrated that the normalized vaporization flux reaches a maximum at the recalescing surface around the icing front and is ~10X larger than that at the supercooled water surface (Figure 3b), i.e., $J_i \sim 10 J_w$, where $J_i$ is the mass flux at the recalescing surface, and $J_w$ is the mass flux at the supercooled droplet surface. See Section S5, Supporting Information for detailed calculations of vaporization fluxes. The non-uniform vaporization results from the temperature and vapor pressure difference at the icing and supercooled liquid water surfaces,[30] and the vaporization flux contrast represents a pulse vapor flow as icing proceeds. Indeed, the strong vapor pulse during recalescence was confirmed by a flow indicator placed next to the icing droplet (Figure S4, Section S3, Supporting Information), consistent with previous observations.[30]

The observed strong progressive vaporization implies counteractive momentum acting on the icing surface, termed vaporization momentum here, which is believed to be responsible for the compressive deformation and jumping directionality of icing droplets on superhydrophobic surfaces. As a direct confirmation of the vaporization momentum, we performed droplet icing on a slippery surface having a low contact angle hysteresis ($\theta_a^{app} - \theta_r^{app} < 8.5°$).[34] The use of slippery surfaces allows us to eliminate surface structuring and superhydrophobicity which may be coupled with the droplet icing and jumping dynamics as is the case with overpressure-induced droplet jumping.[28] Upon depressurization-induced icing, the droplet slides on the surface in the direction along icing propagation (Figure S5, Section S3, Supporting Information). The observed droplet sliding on slippery surfaces along with the droplet deformation on superhydrophobic surfaces indicate the universality of the vaporization momentum of icing droplets at low pressures. Similar to the vaporization momentum mechanism proposed here, evaporation momentum has been also identified in boiling where fast evaporation of heated liquid deforms growing bubbles.[35]

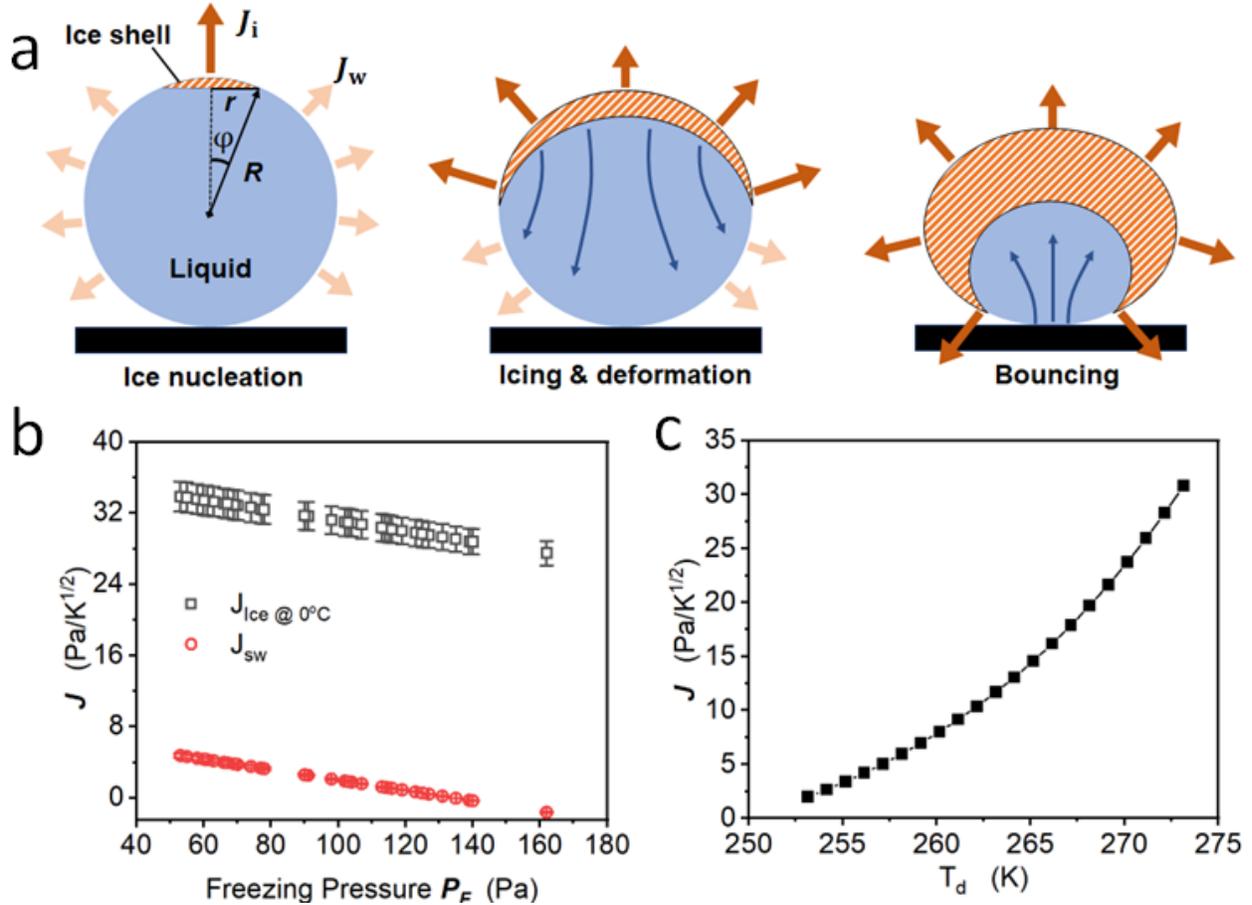

**Figure 3**. Progressive vaporization of the icing droplet. (a) Schematic showing the cross-section of the icing droplet. Ice shell (orange hatched) forms and propagates over the free surface of the droplet. The recalescence elevates the local temperature around the icing front and leads to a higher vaporization flux ($J_i$, red arrows) than that of the supercooled droplet surface ($J_w$, light orange arrows). The flow field within the enclosed water space (light blue) is indicated by dark blue arrows. The length of the arrows indicates the intensity of vaporization (not to scale). As icing proceeds, the droplet is deformed due to the propulsion effects induced by asymmetric vaporization. The substrate counteraction towards the confined liquid redirects the momentum to enable droplet bouncing. (b) The normalized vaporization flux at recalescent ice and supercooled water (sw) surfaces for different environmental pressures. (c) The normalized vaporization flux for different temperature of supercooled droplet

Having identified the role of vaporization momentum in droplet deformation, we then consider if the resulting momentum contributes to droplet jumping. We first analyzed the time scale of droplet detachment $\Delta t_d$, defined as the time duration starting from ice nucleation to droplet-substrate separation. Visualization of droplet icing suggests that the ice shell almost encloses the droplet at the droplet detachment moment, though with a small visible portion remains unfrozen upon detachment (Figure 2). Droplet icing experiments on superhydrophobic surfaces having various microstructures (Figure 4a) show $\Delta t_d$ scales with the inertial-capillary time[41,42] $\tau_c = \sqrt{\rho_w R_d^3/\sigma}$ with a prefactor of ≈2.2 (Figure 4b, see Section S7, Supporting Information), where $\rho_w$ is the density of water, $R_d$ is droplet spherical radius, and $\sigma$ is the surface tension of water. The detaching time ($\Delta t_d \approx 2.2\tau_c$) coincides with the contact time of a low-deformation impact droplet on superhydrophobic surfaces ($\pi/\sqrt{2} \approx 2.2$),[43] suggesting the droplet deformation and detaching dynamics demonstrated here are also likely to be dominated by the inertia and capillary forces of the liquid core, regardless of the solidifying ice shell that gradually encloses the liquid domain of the droplet.

Given the unfrozen liquid core inside the icing droplet, we assume the icing droplet deformation and jumping an elastic droplet bouncing process,[40] where the majority of the vaporization momentum is transferred through the ice shell to the droplet bouncing momentum as the compressive liquid portion elastically interacts with the superhydrophobic substrate. The deformation-bouncing dynamics associated with the liquid core allow us to estimate the possible maximum contribution of vaporization momentum to droplet jumping via the momentum equation:

$$\int_0^{t_d} F_v \cdot dt = \overline{F}_v \Delta t_d = m_d v_j, \qquad (1)$$

where $F_v$ is the vaporization propulsion force exerted on the recalescing surface, $\overline{F}_v$ is the average vaporization propulsion force over the detachment time $\Delta t_d$, $m_d$ and $v_j$ are the mass and translational velocity of the jumping ice droplet, respectively. Given the limited mass loss due to vaporization during the short freezing time, $m_d \approx \rho_w V_d$, where $\rho_w$ and $V_d$ (~$R_d^3$) are the density and volume of the initial water droplet, respectively. Note that gravity is neglected given the droplet' Bond number (Bo = 0.08 – 0.18 < 1).

The vaporization propulsion force scales with the released gas velocity and mass flux, i.e., $F_v \sim J_i A_r v_i$, where the vaporization mass flux $J_i$ of the icing surface is constant for a given system and local vapor pressure.[32,33] $v_i$ is the ejecting velocity of the vapor and $A_r$ (~$R_d^2$) is the surface area of the recalescence surface. Here we ignore the momentum contributed by the vaporization of supercooled liquid due to the significantly lower vaporization mass flux compared to the recalescing surface ($J_i \gg J_w$, Figure 3b). Noting that $\Delta t_d \approx 2.2\tau_c \sim R_d^{1.5}$, Equation 1 is then simplified to $v_j \sim \sqrt{R_d}$ (see Section S7, Supporting Information for detailed derivation). We note that this scaling relationship only represents the upper limit of the jumping velocity as the solidification is assumed not to reduce the bouncing momentum.

To experimentally characterize droplet jumping, we performed droplet (6 to 30 μL) icing experiments on superhydrophobic surfaces with varying microstructures (labeled as S1, S2, and S3 in Figure 4a), characterized by a solid fraction with respect to microstructures ranging from 0.233 to 0.447 and microstructural height ranging from 15 μm to 30 μm. All microstructures are conformally covered by nanostructures having a solid fraction estimated to be 0.108[36] (see Section S2, Supporting information) and a characteristic height of 100 nm.[37] Surfaces having differing microstructures are specially selected to vary the local overpressure underneath the icing droplet, which is shown by previous studies to govern icing droplet self-jumping.[30] The high-speed droplet jumping process was captured to extract the jumping velocity and direction by tracing the droplet trajectory (see Section S4, Supporting Information).[38,39]

Figure 4c shows the experimentally obtained jumping velocity as a function of the root of the droplet radius, showing a linear trend as revealed by the scaling $v_j \sim \sqrt{R_d}$. The scaling analysis shows that a larger droplet size leads to a higher ice jumping velocity, given that the gravity can be ignored under a low Bond number (<1) where surface tension dominates over gravity forces.

So far, the momentum analysis above only considers the progressive vaporization and does not exclude the role of local overpressure in droplet jumping. To further justify our hypothesis that the vaporization momentum dominates over the local overpressure, we calculated the surface structure-dependent overpressure below the freezing droplet based on the balance of local vaporization and vapor drainage:

$$\Delta P = \frac{J_i R_B^2}{\left(\frac{2h}{3(\Delta P + P_{am})}\sqrt{\frac{3R_g T}{\pi M}} + \frac{1}{12\mu}FH^3\right)4\rho_v}, \quad (2)$$

where $R_B$ is the contact radius of the droplet, $H$ is the height of the surface structure, $T$ is the local temperature and is assumed to be the icing surface temperature $T_d$, $P_{am}$ is the ambient pressure, $R_\mathrm{g}$ is the universal gas constant, $\rho_v$ is the density of the vapor, $\mu$ is the viscosity of the vapor, and $M$ is the molecular mass of water vapor. We note that the mass flux released from the recalescing surface ($J_\mathrm{i}$) is used since the droplet detaching happens only after the icing front approaches the droplet base. Equation 2 is a modified version of the semi-analytical correlation given by Ref. 13 to incorporate the slip flow and Knudsen diffusion effects associated with both micro and nanostructures[44] (see Section S6, Supporting Information) while the original version fails when the surface structure length scale reduces to nanoscale. Our refined model addresses this limit by coupling the local pressure/temperature and local vaporization flux to the surface structures, and thus can consistently calculate the overpressure contributed by surface structures having different length scales. The jumping velocity is obtained using the momentum equation having a similar form to Equation 1 by taking the calculated local overpressure as the average overpressure underneath the droplet given the short timescale of the development of the overpressure [13] (see Section S6, Supporting Information for detailed derivation). To validate our overpressure model, we applied Equation 2 to the previously reported experiments[13] and good agreement was demonstrated (Section S4, Supporting Information).

The droplet detaching velocity calculated by the modified overpressure (Equation 2) is shown to underestimate the droplet jumping velocity in our experiments (see Figure S6, Section S6, Supporting Information). More importantly, the local overpressure-induced jumping velocity is predicted to be highly sensitive to substrate structure length scales, and a smaller jumping velocity is gained for surfaces having smaller solid fractions (Figure S6). However, we observe little dependence on jumping velocity on surface micropatterns (pillars or grooves) and solid fractions ($\varphi_\mathrm{micro}$ ranging from 0.22 to 0.447) (Figure 4c). The discrepancy between the overpressure theory and experiments suggests the limited role played by the local overpressure in droplet jumping in our experiments.

To further compare the contribution of the progressive vaporization and local overpressure, we quantified the vaporization momentum in contrast to the overpressure momentum and showed that the maximum vaporization momentum (Equation 1) is 2-3X higher than the overpressure momentum (derived from Equation 2) when the ambient pressure at the freezing moment is below 500 Pa (see Figure S6, Section S6, Supporting Information). This justifies the hypothesis when performing the scaling analysis based on the vaporization-deformation-bouncing dynamics ($v_j \sim \sqrt{R_d}$), where the vaporization momentum dominates over the local overpressure. However, we anticipate that the local overpressure to dominate for ambient pressure over 1000 Pa as is the case with previous studies. [13]

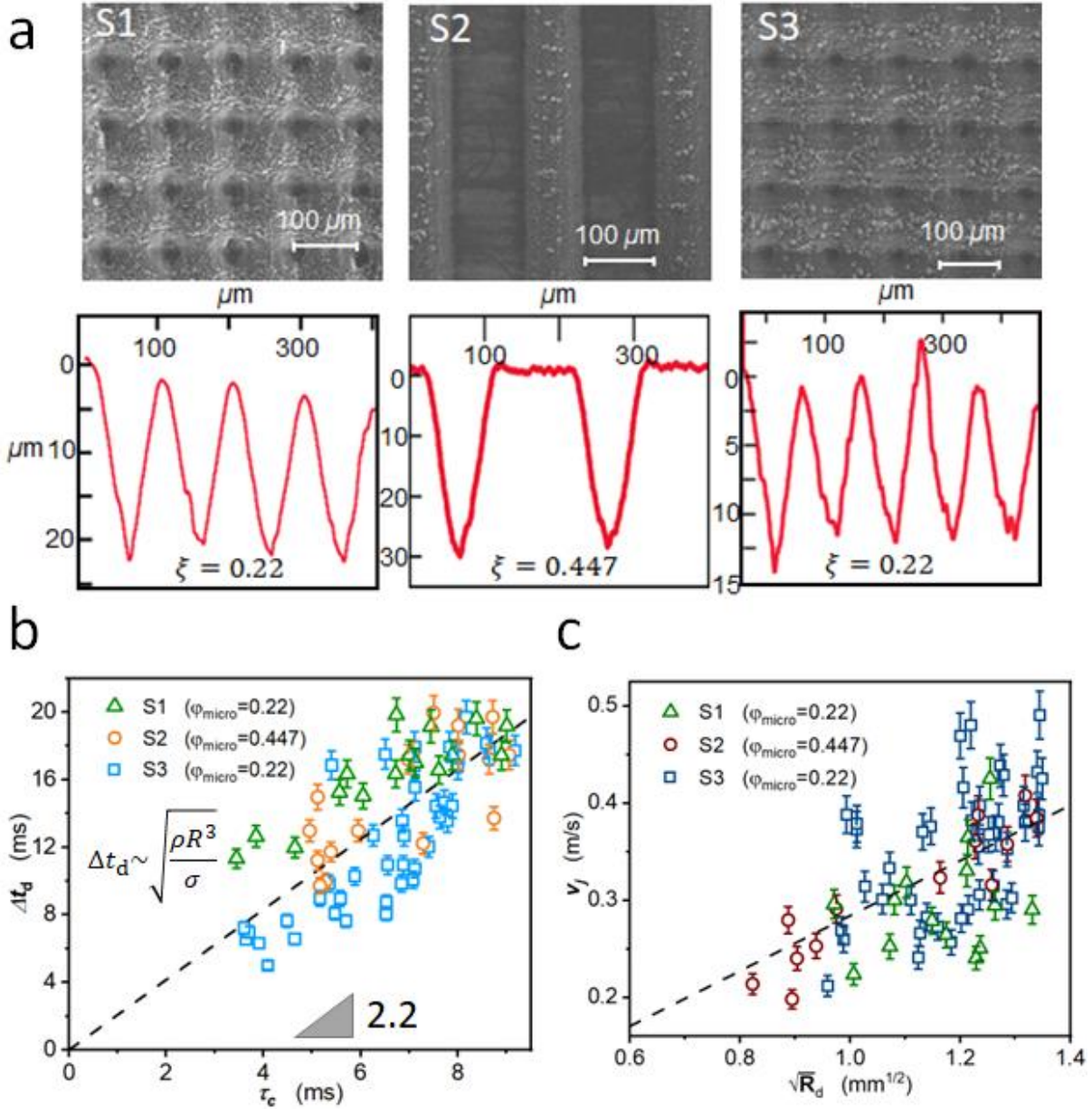

**Figure 4.** (a) Scanning electron microscopy (SEM) of micro/nanostructured substrates, labeled S1, S2, and S3, along with their cross-section profiles below the SEM images. Solid fractions associated with microstructures ($\varphi_{micro}$) are presented in the cross-section profiles. (b) Detachment time $\Delta t_d$ as a function of capillary time $\tau_c$ for droplets having various spherical radii ($R_d$, measured immediately before icing) on different substrates. The dotted line represents the linear fitting with a slope of 2.2. (c) Droplet jumping velocity $v_j$ as a function of the square-rooted droplet spherical radius $\sqrt{R_d}$ for different substrates. See Section S2 for more details on the substrates.

The vaporization momentum is further shown to govern the jumping direction. Since the initial nucleation site represents the start of progressive recalescence, it determines the spatial distribution of vaporization fluxes over the droplet surface and thus the jumping direction. We define the jumping direction by the jumping angle ($\theta_j$) with respect to the substrate normal line at the detaching moment, which is obtained by the jumping trajectory analysis[31] (see Section S4, Supporting information). The angular position ($\theta_N$) of the initial ice nucleation site with respect to the droplet center of mass (COM) is used to characterize the spatial distribution of ice nucleation (Figure 5a). It was shown that $\theta_j$ is linearly correlated with $\omega_N$ independent of droplet sizes and substrates, i.e., $\theta_j = -\theta_N$ (Figure 5b). Figure 5c demonstrates a droplet with an ice nucleation initiation position on the right-top interface of the droplet ($\theta_N = -34.2°$), while the icing droplet jumps towards left-top (e.g., $\theta_j = 30.3°$). Consistent with our momentum transfer analysis above, the vaporization momentum is redirected to form the jumping momentum as the liquid phase of the droplet interacts with the substrate in an elastic bouncing-like manner.[39,45] As a result, the jumping direction is determined by the reflection of the vaporization momentum that pushes the droplet to deform. The jumping directionality observed here, which cannot be explained by the overpressure mechanism,[13] thus further highlights the significant role of vaporization momentum in icing droplet jumping.

The ice nucleation position and therefore the jumping direction is experimentally observed to be probabilistic. By statistically sampling the data from repeated experiments, we found that the distribution of jumping direction (and ice nucleation position), quantified by the relative frequency $\phi$, follows the Gaussian density function (Figure 5d), which reveals that the droplet is more probable to jump perpendicularly to the substrate ($\theta_j \approx 0°$) and less probable to jump in plane ($\theta_j \approx \pm 90°$). This is because ice nucleation tends to initiate at the top of the supercooled droplet, where the temperature is the lowest according to the evaporation heat transfer prediction.[32,33] Furthermore, the symmetry of the jumping direction distribution suggests that the icing droplet holds an equal probability of jumping leftwards and rightwards, indicative of the minimal effects of experimental conditions (e.g., airflow during vacuuming) on droplet icing and jumping.

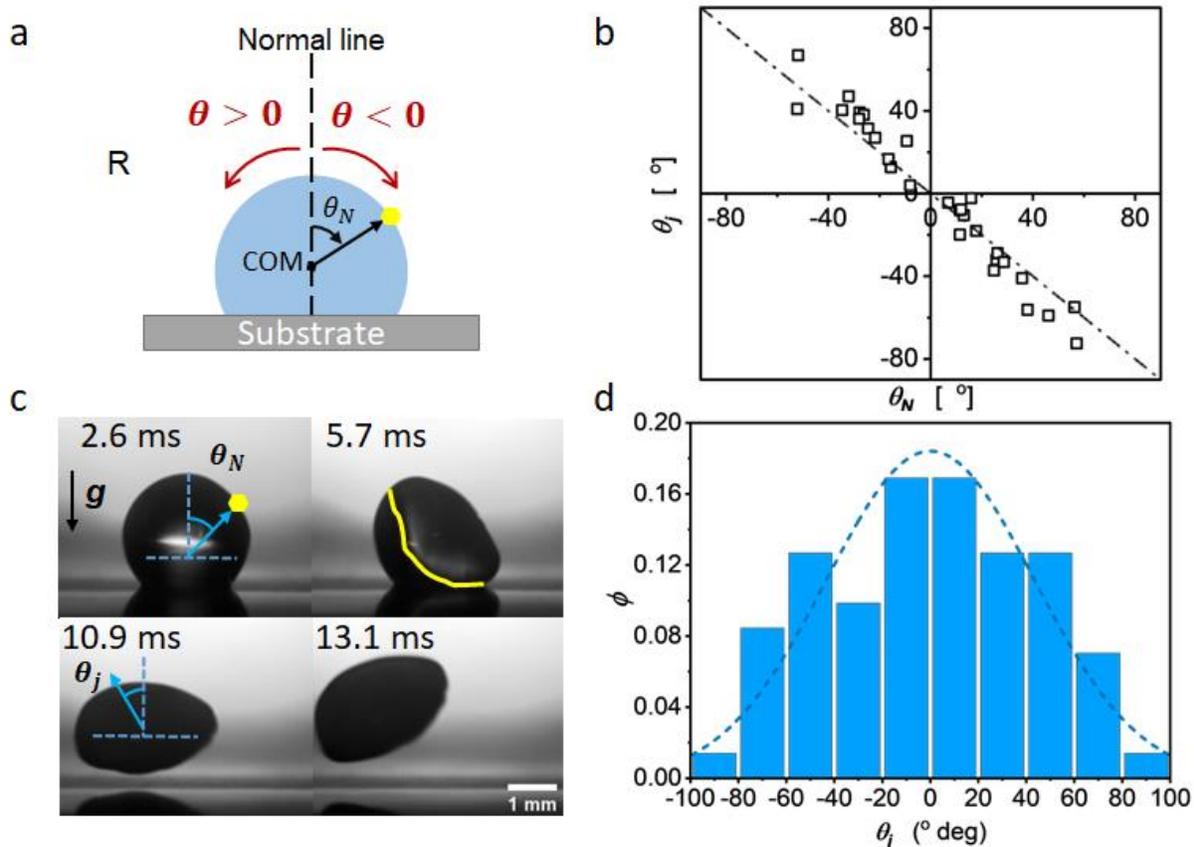

**Figure 5.** (a) Schematic showing the definition of jumping angle ($\theta_j$) and the angular position of the nucleation site ($\theta_N$). The normal line with respect to the substrate represents $\theta_j=0°$ or $\theta_N=0°$. An angle towards to right is defined to be positive and vice versa. (b) Jumping angle $\theta_j$ as a function of the angular position of the ice nucleation site $\theta_N$, showing a linear trend, i.e., $\theta_j=-\theta_N$. (c) An example demonstrating the relationship of jumping angle $\theta_j$ and angular position of the nucleation site $\theta_N$. The nucleation site is marked by a hexagon. (d) The relative frequency $\phi$ of the jumping angle $\theta_j$, which is fitted using a Gaussian density function, $\phi = \frac{1}{43.3\sqrt{2\pi}} e^{-0.5\left(\frac{\theta_j}{43.3}\right)^2}$, with a mean at ≈0° and a standard deviation of ≈43°, represented by the dotted curve.

In conclusion, we have demonstrated the physics governing supercooled water droplet icing and jumping from superhydrophobic surfaces at low pressures (~100 Pa). Although observed at a high depressurization rate (100 kPa to 0.1 kPa within ~10 seconds), the icing and droplet self-detaching phenomena indicate a universal mechanism with respect to droplet and surface interaction coupled with vaporization momentum. Different from the previously reported local overpressure underneath the droplet that drives droplet trampolining, the vaporization momentum arises from the globally asymmetric vaporization flux and imparts a counteractive force to the free surface of the icing droplet, leading to the compressive deformation of the droplet as ice front propagates. We quantified this vaporization momentum in comparison with the local overpressure and showed that the former dominates at ambient pressures of ~100 Pa. We proposed a vaporization-compression-bouncing physical model to depict the droplet icing and jumping dynamics, which enables us to predict the jumping velocity as a function of droplet size. Furthermore, the physical model also reveals that the spatially probabilistic nature of the ice nucleation site governs the jumping direction, as confirmed by the experimental statistics. At a higher pressure approaching ~1000 Pa, we anticipate the local overpressure mechanism to play a more important role than the vaporization momentum due to the reduced vaporization flux. At such conditions, the coupling between the top-down and bottom-up mechanisms demands further investigation. Our study provides insights into the phase change of supercooled water droplets and presents an approach to detach supercooled droplets from surfaces at decreased pressures. The demonstrated physics may help facilitate water management in vacuum and aerospace-related applications.

## ASSOCIATED CONTENT

**Supporting Information**

The Supporting Information is available free of charge at XXXX.

Fabrication and characterization of the superhydrophobic substrates; Measurement of the droplet freezing temperature; Characterization of ice droplet jumping; Estimation of ice sublimation flux and supercooled water evaporation flux; Quantification of overpressure effect on jumping velocity; Physical model of icing droplet jumping; Calculation of deformation ratio; Estimation of the solid fraction of nanostructure; Ice sliding and rebounding experiment; Characterization of the pressure chamber.

## AUTHOR INFORMATION

**Corresponding Author**

**Shuhuai Yao** – Department of Mechanical and Aerospace Engineering, The Hong Kong University of Science and Technology, Hong Kong 999077, China; orcid.org/0000- 0001-7059-4092; Email: meshyao@ust.hk

**Author Contributions**

X.Y. and S.A. contributed equally to this work. X.Y., S.A., and S.Y. conceived the idea for the work. X.Y. designed and guided the experiments. S.A., X.Y., S.C., Y.C., N.L, and W.W. performed experiments. G.Z., X.Y., and M.M. fabricated samples for experiments. D.S. performed surface characterization. S.A. and X.Y. analyzed the data, performed modeling, and wrote the manuscript. S.Y. supervised the work and edited the manuscript.

**Notes**

The authors declare no competing financial interest.

**ACKNOWLEDGMENTS**

This work was financially supported by the Research Grants Council of Hong Kong under the General Research Fund (16213721). The authors appreciate the help from Dr. Jixiang Wang at Yangzhou University and Prof. Weihong Li at City University of Hong Kong. X.Y. also appreciates Prof. Nenad Miljkovic for his kind support.

**Table of Content**

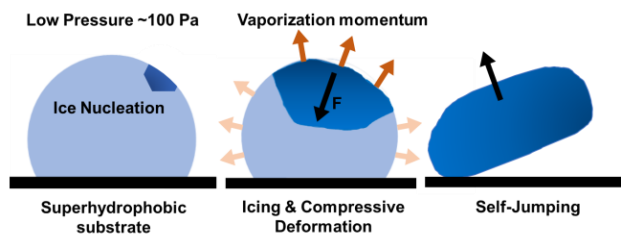